\documentclass[twocolumn,aps,prc,showpacs,floatfix]{revtex4}
\usepackage{graphicx}
\usepackage{amsmath}
\usepackage{dcolumn}
\usepackage{bm}

\begin{document}

\title{Cross-checking the symmetry energy at high densities}

\author{Gao-Chan Yong}

\affiliation{%
{Institute of Modern Physics, Chinese Academy of Sciences, Lanzhou
730000, China}
}%

\begin{abstract}
By considering both the effects of the
nucleon-nucleon short-range correlations and the isospin-dependent
in-medium inelastic baryon-baryon scattering cross section in the
transport model, two unrelated Au + Au experimental measurements at 400 MeV/nucleon beam energy are simultaneously analyzed, a mildly soft symmetry energy ($L (\rho_{0})$ = 37 MeV) at supra-saturation densities is obtained. This result is compatible with the recent result in Phys. Rev. C {\bf 92}, 064304 (2015) by comparing the available data on the electric dipole polarizability with the theoretical predictions.

\end{abstract}

\maketitle

\section{Introduction}
The nuclear symmetry energy describes the single nucleonic energy
of nuclei or nuclear matter changes as one replaces protons in a
system with neutrons. Besides its impacts in nuclear physics
\cite{Bar05,LCK08}, in a density range of 0.1 $\sim$ 10 times
nuclear saturation density, the symmetry energy determines the
birth of neutron stars and supernova neutrinos \cite{Sumiyoshi95},
a range of neutron star properties such as cooling rates, the
thickness of the crust, the mass-radius relationship, and the
moment of inertia \cite{Sum94,Lat04,Ste05a,Lattimer14}. The
nuclear symmetry energy also plays crucial role in the evolution
of core-collapse supernova \cite{Fischer14} and astrophysical
r-process nucleosynthesis \cite{Nikolov11}. Thus the better we can
constrain the symmetry energy in laboratory measurements, the more
we can learn from astro-observations.

To constrain the symmetry energy in broad density regions, besides
the studies in astrophysics \cite{apj10,fatt13,fatt14}, many terrestrial
experiments are being carried out or planned using a wide variety
of advanced new facilities, such as the Facility for Rare Isotope
Beams (FRIB) in the US \cite{frib}, or the Radioactive Isotope Beam Facility
(RIBF) in Japan \cite{ribf}. To unscramble symmetry energy related
experimental data, various isospin-dependent transport models are
frequently used to probe the symmetry energy below and above
saturation density
\cite{LCK08,Bar05}. With great efforts, the nuclear symmetry
energy and its slope around saturation density of nuclear matter
from 28 analysis of terrestrial nuclear laboratory experiments and
astrophysical observations have been roughly pinned down
\cite{lihan13}, while recent interpretations of the FOPI and FOPI-LAND experimental
measurements by different groups made the symmetry energy at
supra-saturation densities fall into chaos \cite{isog2015,xiao09,fengplb,russplb,cozma13,xieplb,wangprc}.
It does not seem to be clarified why the nuclear symmetry energy at supra-saturation densities is so uncertain, maybe the effects of pion in-medium effects \cite{WMGuo15,hongj2014,xuj2013}, the isospin dependence of in-medium nuclear strong interactions \cite{yong2011}, the short-range tensor force \cite{VRP72,xuc2011} are some factors.

Recently, the high-momentum transfer measurements showed that
nucleons in nucleus can form pairs with large relative momenta
and small center-of-mass momenta \cite{pia06,sh07}. This
phenomenon was explained by the short-range nucleon-nucleon tensor
interaction \cite{tenf05,tenf07}. Such nucleon-nucleon short-range
correlations (SRC) in nucleus lead to a high-momentum tail (HMT)
in the single-nucleon momentum distribution \cite{bethe71,anto88,Rios09,yin13}.
More interestingly, in the HMT of nucleon momentum distribution,
nucleon component is evidently isospin-dependent. The number of
n-p SRC pairs is about 18 times that of the p-p and n-n SRC pairs
\cite{sci08}. And in neutron-rich nucleus, proton has a greater
probability than neutron to have momentum greater than the
nuclear Fermi momentum \cite{sci14}.

Unfortunately, effects of the above isospin-dependent SRC were
seldom taken into account in most of currently used
isospin-dependent transport models, while the latters have been
frequently used to unscramble symmetry energy related
experimental data
\cite{Shetty07,betty09,lynch09,xiao09,cozma13,natowitz10}. To extract
information on the symmetry energy from experimental data, in this
study, by considering both the effects of the isospin-dependent SRC and
the important but often-overlooked in-medium baryon-baryon
inelastic cross section in the isospin-dependent transport model,
two unrelated experimental measurements are simultaneously
re-analyzed.

\section{The IBUU transport model}
\begin{figure}[th]
\centering
\includegraphics[width=0.5\textwidth]{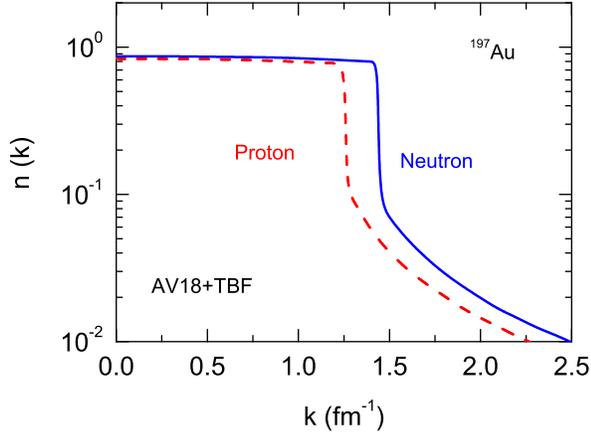}
\caption{(Color online) Momentum distributions of neutron and proton in nucleus $^{197}$Au calculated with
the BHF with Av18+TBF. Taken from Ref.~\cite{yong2015}.} \label{npdis}
\end{figure}
\begin{figure}[th]
\centering
\includegraphics[width=0.5\textwidth]{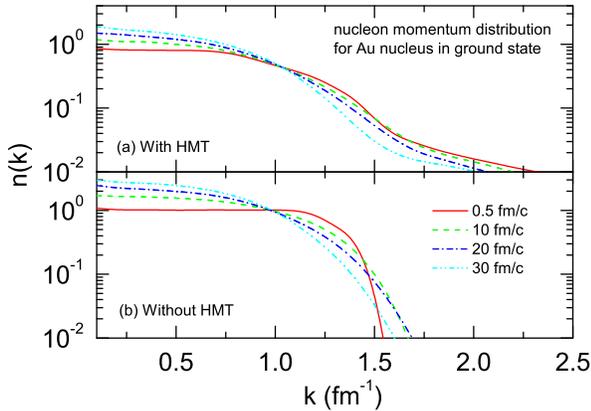}
\caption{(Color online) Evolution of the momentum distribution of nucleon in nucleus $^{197}$Au. The upper panel (a) is for the case of initial nucleon distribution with HMT while the lower panel (b) is a simple
Fermi-Dirac initial distribution. Both cases are under the interaction given by Eq.~(\ref{buupotential}).} \label{mdis}
\end{figure}
\begin{figure}[th]
\centering
\includegraphics[width=0.5\textwidth]{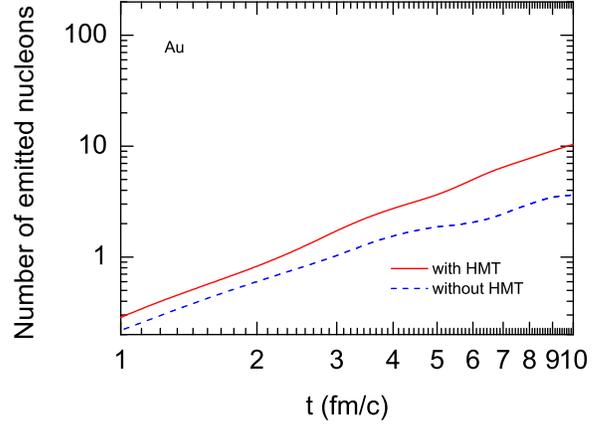}
\caption{(Color online) Time dependence of the number of evaporated nucleons by
the nucleus in ground state using two initial distributions with and without high momentum tail as shown in Fig.~\ref{mdis}.} \label{nemit}
\end{figure}
\begin{figure}[th]
\centering
\includegraphics[width=0.5\textwidth]{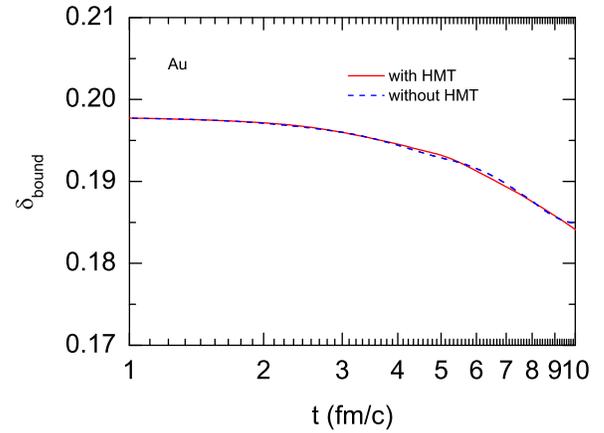}
\caption{(Color online) Time dependence of the asymmetry $\delta_{bound} = (\rho_n-\rho_p)/(\rho_n+\rho_p)$ of bound nucleons in
the nucleus in ground state using two initial distributions with and without high momentum tail as shown in Fig.~\ref{mdis} and Fig.~\ref{nemit}.} \label{deltat}
\end{figure}

\begin{figure}[th]
\centering
\includegraphics[width=0.5\textwidth]{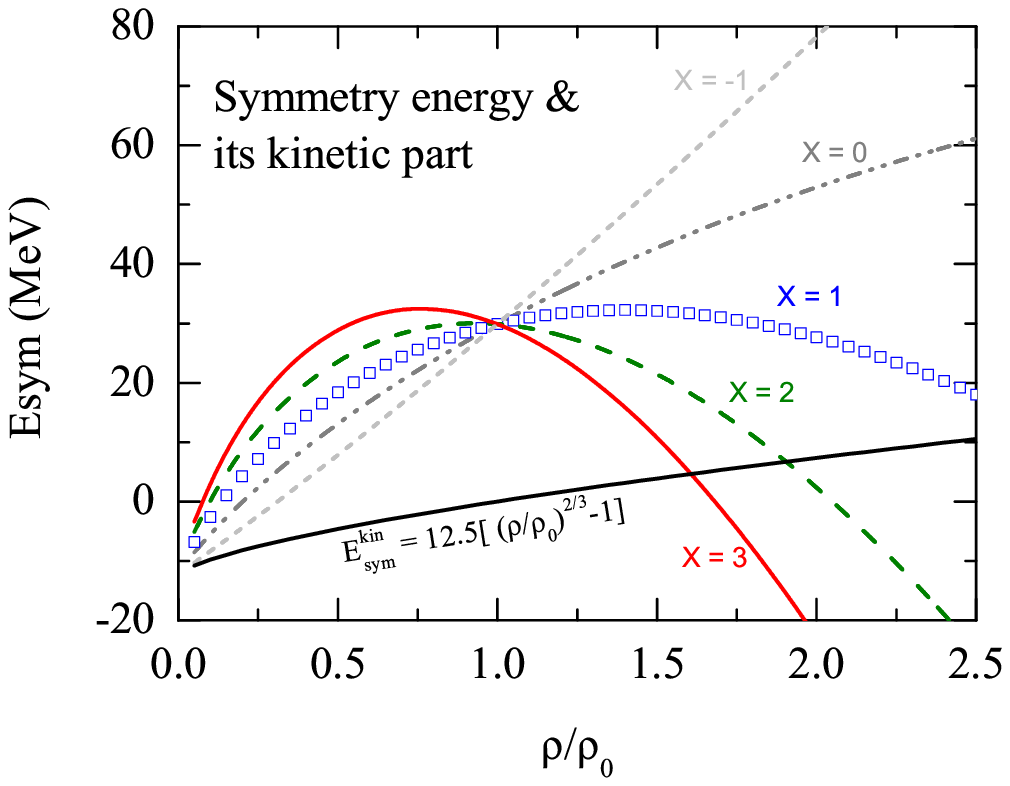}
\caption{(Color online) Kinetic symmetry energy and density-dependent symmetry energy with
different $x$ parameters. Note here that in heavy-ion collisions at 400 MeV/nucleon beam energy, the high-density part of the symmetry energy plays major role.} \label{esym}
\end{figure}
To probe the symmetry energy from experimental data,
we use our recent updated Isospin-dependent
Boltzmann-Uehling-Uhlenbeck (IBUU) transport model \cite{yong2015}.
In this IBUU model, nucleon-density
distribution is given by
\begin{equation*}
r = R(x_{1})^{1/3}; cos\theta = 1-2x_{2}; \phi = 2\pi x_{3};
\end{equation*}
\begin{equation}\label{xyz}
    x = rsin\theta cos\phi;
    y = rsin\theta sin\phi;
    z = rcos\theta.
\end{equation}
Where $R$ is the radius of nucleus, $x_{1}, x_{2}, x_{3}$ are
three independent random numbers.
Since there is a depletion of nucleon distribution inside the Fermi sea, the proton and neutron
momentum distributions with high-momentum tail reaching about 2 times the Fermi momentum \cite{yong2015} are given by the extended Brueckner-Hartree-Fock (BHF)
approach by adopting the AV 18 two-body interaction plus a microscopic Three-Body-Force (TBF) \cite{yin13}. Fig.~\ref{npdis} shows such nucleon momentum distribution with high-momentum tail in $^{197}$Au. Compared with the distribution in ideal Fermi gas, the excess energy of nucleon in colliding nuclei is subtracted from the total energy of reaction system.

How long will survive the shape of the initial
distributions? To answer this question, Fig.~\ref{mdis} shows the
plots of the implemented distributions at several times
(0.5 fm/c, 10 fm/c, 20 fm/c, 30 fm/c) for nuclei in the ground states when
they are left to evolve under the interaction given by
Eq.~(\ref{buupotential}) in this text and initialized in agreement with the distribution seen
in Fig.~\ref{npdis}. The lower panel (b) shown in Fig.~\ref{mdis} is under the simple
Fermi-Dirac distribution as comparison. It is seen that the shape of the initial
distribution of nucleon in momentum space is relatively well kept at the initial stage of collision at 400 MeV/nucleon incident beam energy.
Related to the previous point, it is important to show a comparative
study of the time dependence of the number of evaporated nucleons by
the nuclei in ground state, both when are considered a simple
Fermi-Dirac distribution and a distribution with high momentum tail.
Shown in Fig.~\ref{nemit} is the time dependence of the number of evaporated nucleons by
the nucleus in ground state using two initial distributions with and without high momentum tail. One can clearly see that compared with the case without HMT, more nucleons emit from nucleus with HMT at the initial stage of collision at 400 MeV/nucleon incident beam energy. However, to study the effect of the symmetry energy, the asymmetry ($\delta_{bound} = (\rho_n-\rho_p)/(\rho_n+\rho_p)$) of the colliding system is more crucial. Shown in Fig.~\ref{deltat} is the time dependence of the asymmetry of bound nucleons in
the nucleus in ground state using two initial distributions with and without high momentum tail as shown in Fig.~\ref{mdis} and Fig.~\ref{nemit}. It is seen that with HMT the asymmetry of colliding nucleons in the reaction system is almost the same as that without HMT at the the initial stage of collision. One can thus conclude that although the present HMT considerations in initial colliding nuclei cause the instability of nuclei, it in fact does not affect much the present study of the symmetry energy at high densities using heavy-ion collisions.

In this model, the isospin- and momentum-dependent mean-field
single nucleon potential is used \cite{Das03,xu14,yong2015}, i.e.,
\begin{eqnarray}
U(\rho,\delta,\vec{p},\tau)&=&A_u(x)\frac{\rho_{\tau'}}{\rho_0}+A_l(x)\frac{\rho_{\tau}}{\rho_0}\nonumber\\
& &+B(\frac{\rho}{\rho_0})^{\sigma}(1-x\delta^2)-8x\tau\frac{B}{\sigma+1}\frac{\rho^{\sigma-1}}{\rho_0^\sigma}\delta\rho_{\tau'}\nonumber\\
& &+\frac{2C_{\tau,\tau}}{\rho_0}\int
d^3\,\vec{p^{'}}\frac{f_\tau(\vec{r},\vec{p^{'}})}{1+(\vec{p}-\vec{p^{'}})^2/\Lambda^2}\nonumber\\
& &+\frac{2C_{\tau,\tau'}}{\rho_0}\int
d^3\,\vec{p^{'}}\frac{f_{\tau'}(\vec{r},\vec{p^{'}})}{1+(\vec{p}-\vec{p^{'}})^2/\Lambda^2},
\label{buupotential}
\end{eqnarray}
where $\tau, \tau'=1/2(-1/2)$ for neutrons (protons),
$\delta=(\rho_n-\rho_p)/(\rho_n+\rho_p)$ is the isospin asymmetry,
and $\rho_n$, $\rho_p$ denote neutron and proton densities,
respectively. Specifically, the parameter values $A_u(x)$ = 33.037 - 125.34$x$
MeV, $A_l(x)$ = -166.963 + 125.34$x$ MeV, B = 141.96 MeV,
$C_{\tau,\tau}$ = 18.177 MeV, $C_{\tau,\tau'}$ = -178.365 MeV $\sigma =
1.265$, and $\Lambda = 630.24$ MeV/c are obtained by fitting seven
empirical constraints of the saturation density $\rho_{0}$ = 0.16
fm$^{-3}$, the binding energy $E_{0}$ = -16 MeV, the
incompressibility $K_{0}$ = 230 MeV, the isoscalar effective mass
$m_{s}^{*} = 0.7 m$, the single-particle potential
$U^{0}_{\infty}$ = 75 MeV at infinitely large nucleon momentum at
saturation density in symmetric nuclear matter, the symmetry
energy $S(\rho)$ = 30 MeV and the symmetry potential
$U^{sym}_{\infty}$ = -100 MeV at infinitely large nucleon momentum
at saturation density. $f_{\tau}(\vec{r},\vec{p})$ is
the phase-space distribution function at coordinate $\vec{r}$ and
momentum $\vec{p}$ and solved by using the test-particle method
numerically \cite{yong2015}.
Different symmetry energy's stiffness parameter $x$
can be used in the above single nucleon potential to mimic
different forms of the symmetry energy. Since the kinetic symmetry energy, even its sign, is
still controversial \cite{henprc15}, we at present give it a null value \cite{Isaac2011}. For its
density-dependence, we use similar form as that from the ideal Fermi gas model. Thus the density-dependent kinetic symmetry energy is expressed as
\begin{eqnarray}
E_{sym}^{kin} = 12.5 [(\rho/\rho_{0})^{2/3}-1].
\end{eqnarray}
Fig.~\ref{esym} shows the kinetic symmetry energy we used and the density-dependent symmetry energy with different $x$ parameters. It is seen that our used density-dependent kinetic symmetry energy is similar to that in Ref. \cite{henprc15,Carbone2012}.
We can also see that $x = 1, 0, -1 $ cases roughly correspond positive slopes ($L (\rho_{0}) \equiv 3\rho_{0}dEsym(\rho)/d\rho$) 37, 87, 138 MeV, respectively.
The following table shows the parameters used in Eq.~(\ref{buupotential}) with SRC compared with the case without SRC \cite{Das03}.
\begin{table}[th]
\caption{The parameters (in MeV) used in Eq.~(\ref{buupotential}) with SRC compared with the case without SRC \cite{Das03}.}
\label{notef3}%
\begin{tabular}{|c|c|c|}
  \hline
  $E_{sym}^{kin}$  & 0 ( = with SRC) & 12.5 $(\rho/\rho_{0})^{2/3}$ ( = without SRC)\\
  \hline
  $A_u(x)$         & 33.037 - 125.34$x$ & -95.98 - 91.157$x$ \\
  \hline
  $A_l(x)$         & -166.963 + 125.34$x$ & -120.57 + 91.157$x$ \\
  \hline
  $B$              & 141.96 & 106.35 \\
  \hline
  $C_{\tau,\tau}$  & 18.177 & -11.7 \\
  \hline
  $C_{\tau,\tau'}$ & -178.365 & -103.4 \\
  \hline
  $\sigma$         & 1.265 & 1.333 \\
  \hline
  $\Lambda$        & 630.24 & 260 \\
  \hline
\end{tabular}
\end{table}

\begin{figure}[th]
\centering
\includegraphics[width=0.5\textwidth]{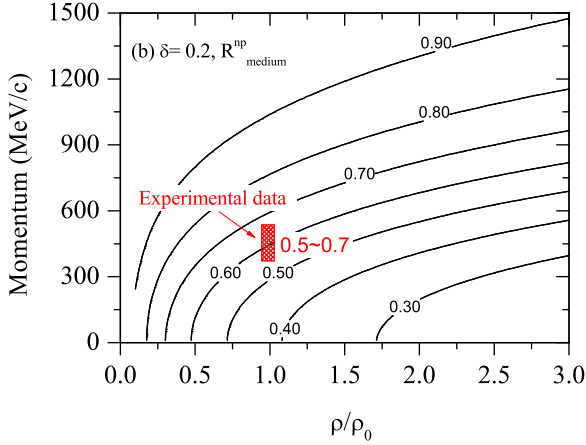}
\caption{(Color online) Reduced factor $R^{np}_{medium}$ of neutron-proton
scattering cross section as a function of density and nucleonic
momentum in asymmetric medium $\delta$ = 0.2.
The red shadow region and label `0.5$\sim$0.7' denote experimental
density and momentum region and the corresponding value of reduced
factor $R^{np}_{medium}$.} \label{npx}
\end{figure}
The isospin-dependent baryon-baryon ($BB$) scattering cross section (elastic or inelastic, including NN$\rightarrow$NN, N$\Delta$$\rightarrow$N$\Delta$, $\Delta$$\Delta$$\rightarrow$$\Delta$$\Delta$, NN$\rightleftharpoons$N$\Delta$) in medium $\sigma
_{BB}^{medium}$ is reduced compared with their free-space value
$\sigma _{BB}^{free}$ by a factor of \cite{yong2015}
\begin{eqnarray}
R^{BB}_{medium}(\rho,\delta,\vec{p})&\equiv& \sigma
_{BB_{elastic, inelastic}}^{medium}/\sigma
_{BB_{elastic, inelastic}}^{free}\nonumber\\
&=&(\mu _{BB}^{\ast }/\mu _{BB})^{2},
\end{eqnarray}
where $\mu _{BB}$ and $\mu _{BB}^{\ast }$ are the reduced masses
of the colliding baryon-pair in free space and medium (in medium, the effective mass of baryon  is used),
respectively. The effective mass of baryon in isospin asymmetric nuclear matter
is expressed by
\begin{equation}
\frac{m_{B}^{\ast }}{m_{B}}=\left\{ 1+\frac{m_{B}}{p}\frac{%
dU_{B}}{dp}\right\}.
\end{equation}
For the baryon resonance $\Delta$ potential, the forms of
\begin{eqnarray}
U^{\Delta^-}_{B}&=& U_{n},\\
U^{\Delta^0}_{B}&=& \frac{2}{3}U_{n}+\frac{1}{3}U_{p},\\
U^{\Delta^+}_{B}&=& \frac{1}{3}U_{n}+\frac{2}{3}U_{p},\\
U^{\Delta^{++}}_{B}&=& U_{p}
\end{eqnarray}
are used \cite{delta15}. As an example, the reduced factor $R^{np}_{medium}(\rho,\delta,\vec{p})$
of neutron-proton scattering cross section in medium is shown in
Fig.~\ref{npx}.
It is a function of
density and nucleonic momentum in medium with asymmetry $\delta= 0.2$ for
example. And generally $R^{nn}_{medium}
> R^{np}_{medium} > R^{pp}_{medium}$ at certain density and
momentum. The nucleon-nucleon scattering cross section is
reduced much in medium for colliding pair at high density and low
momentum while it is less reduced at low density and high
momentum.
It is noticed from Fig.~\ref{npx} that the used
in-medium neutron-proton scattering cross section fits the
experimental hard photon measurements quite well \cite{yong11}.

\section{Results and discussions}
\begin{figure}[th]
\centering
\includegraphics[width=0.5\textwidth]{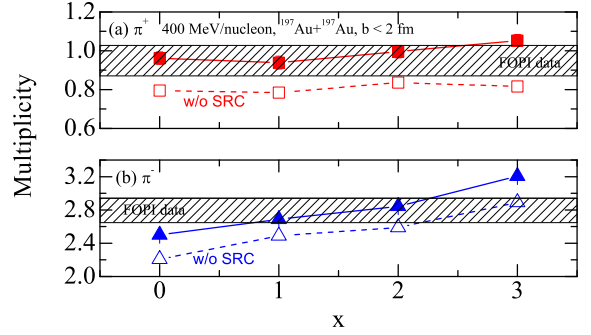}
\caption{(Color online) Multiplicity of charged pion meson produced in Au+Au reaction at 400 MeV/nucleon with different symmetry energies. The shadow region denotes the FOPI data \cite{Reisdorf10}. The dashed lines show the cases without SRC.} \label{multipion}
\end{figure}
Before studying the $\pi^{-}/\pi^{+}$ ratio, it is instructive to first see the production of charged pion meson in central Au + Au reaction at 400 MeV/nucleon beam energy. Fig.~\ref{multipion} shows numbers of charged pion produced with different symmetry energies.
It is seen that for $x$ = 1 and $x$ = 2 cases both produced $\pi^{-}$ and $\pi^{+}$ fit the FOPI experimental data quite well.
With stiffer symmetry energy $x = 0$, the model gives somewhat smaller $\pi^{-}$ number than experimental data. While with very soft symmetry energy $x = 3$, the model gives both larger $\pi^{+}$ and $\pi^{-}$ numbers than experimental data. Comparing upper panel with lower panel, it is seen that sensitivity of the number of produced $\pi^{-}$ to the symmetry energy is at least 3 times that of $\pi^{+}$. This is because the $\pi^{-}$ mesons are mostly produced from
neutron-neutron collisions, thus more sensitive to the isospin asymmetry
of the reaction system and the symmetry energy \cite{lyz05}.
From this figure, one can see that without SRC, numbers of both $\pi^{+}$ and $\pi^{-}$ are smaller than those with SRC. This is understandable since the SRC of nucleons increases the kinetic energy of nucleon.

\begin{figure}[th]
\centering
\includegraphics[width=0.5\textwidth]{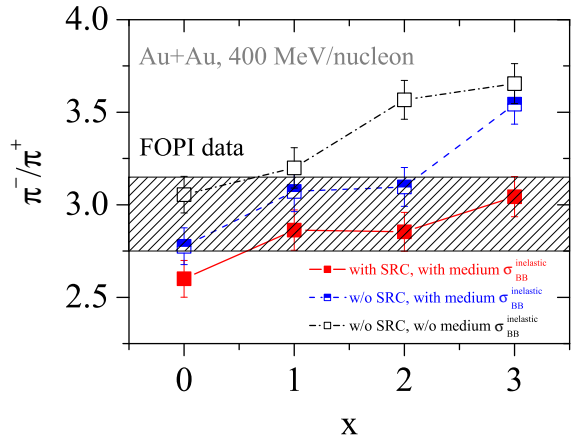}
\caption{(Color online) $\pi^{-}/\pi^{+}$ ratio in Au+Au reaction at 400 MeV/nucleon with different symmetry energies. Also shown are the effects of the SRC of nucleon-nucleon and the in-medium inelastic cross section on the $\pi^{-}/\pi^{+}$ ratio with same $x$ parameters.} \label{rpion}
\end{figure}
To reduce the systematic errors,
most of the observables proposed so far use differences or ratios of isospin multiplets
of baryons, mirror nuclei and mesons, such as, the
neutron/proton ratio of nucleon emissions, neutron-proton
differential flow, $\pi^{-}/\pi^{+}$, etc.
Fig.~\ref{rpion} shows the $\pi^{-}/\pi^{+}$ ratio predicted by our IBUU model with different symmetry energies. Because softer symmetry energy causes more neutron-rich dense matter and $\pi^{-}$'s are mainly from neutron-neutron collision whereas $\pi^{+}$'s are
mainly from proton-proton collision \cite{lyz05}, it is
not surprising that one sees larger $\pi^{-}/\pi^{+}$ ratio with
softer symmetry energy. To see the effects of SRC of nucleon-nucleon and the reduction of the
in-medium inelastic baryon-baryon scattering cross section, with same $x$ parameters, we
made calculations by turning off the SRC and the reduction of the
in-medium inelastic baryon-baryon scattering cross section, respectively. From Fig.~\ref{rpion},
we can see that both of them affect the value of $\pi^{-}/\pi^{+}$ ratio evidently. Both the SRC of nucleon-nucleon and the reduction of the in-medium inelastic baryon-baryon scattering cross section
decrease the value of $\pi^{-}/\pi^{+}$ ratio evidently. Proton-proton collision is also affected by the Coulomb action, so $\pi^{+}$ production, which is mainly from proton-proton collision, is relatively less affected by the reduction of the in-medium inelastic baryon-baryon scattering cross section.
\begin{figure}[th]
\centering
\includegraphics[width=0.5\textwidth]{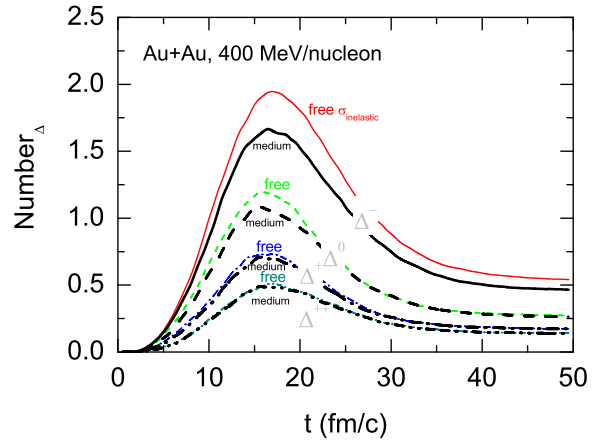}
\caption{(Color online) The effects of in-medium inelastic cross section on the productions of different $\Delta$ resonance in Au+Au reaction at 400 MeV/nucleon with the symmetry energy parameter $x = 1$.} \label{meddelta}
\end{figure}
Fig.~\ref{meddelta} shows the evolution of $\Delta$ resonance production with free and in-medium
inelastic cross sections. It is clear seen that the in-medium inelastic cross section affects the production of $\Delta^{-}$ (decay into $\pi^{-}$) much than $\Delta^{++}$ (decay into $\pi^{+}$). This is the reason why the reduced in-medium inelastic baryon-baryon cross section decreases the value of $\pi^{-}/\pi^{+}$ ratio.
The SRC of neutron and proton causes small asymmetry of matter, which corresponds small value of $\pi^{-}/\pi^{+}$ ratio. Therefore both the SRC of nucleon-nucleon and the in-medium inelastic baryon-baryon cross sextion should be taken into account in transport calculations. From Fig.~\ref{rpion},
we can see that the FOPI pion experimental data supports a softer symmetry energy ($x$ = 1, 2, even $x$ = 3).
Note here that the density region probed here is about $1\sim1.5$ times saturation density \cite{liuy2015}. So we do not conclude what are the constraints of the values of the symmetry energy and
its slope around saturation density.

While $\triangle$ and $\pi$'s scattering and re-absorption can destroy the high density signal in a certain degree. Treatment of Delta dynamics in transport models is not so straightforward such as the competing effects of the mean fields and $\triangle$ thresholds. To understand quantitatively the symmetry energy effect on pion production, it is important to include the isospin-dependent pion in-medium effects \cite{xuj2013,WMGuo15,hongj2014}. And recent work of MSU group \cite{pawel2015} demonstrates that the ratio of $\pi$'s spectra is more sensitive than the ratio of integrated yields because the latter gives ambiguous result since it does not distinguish $\pi$'s messenger of high density from the rest. Therefore, more theoretical and experimental studies are needed to pin down the high-density behavior of the symmetry energy by pion probe.

\begin{figure}[th]
\centering
\includegraphics[width=0.5\textwidth]{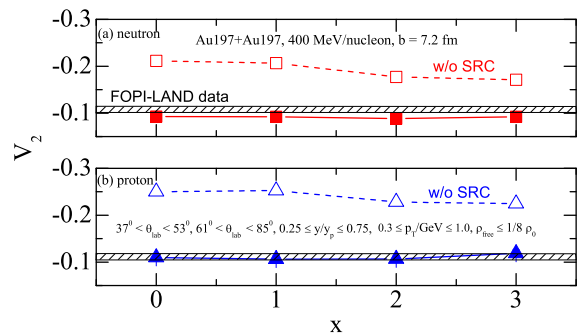}
\caption{(Color online) Elliptic flow of emitting nucleons in Au+Au collision at 400 MeV per
nucleon incident beam energy with different symmetry energies. The
shadow region denotes the experimental FOPI-LAND data \cite{cozma13}. The dashed lines show the cases without SRC.}
\label{v2np}
\end{figure}
To cross-check the symmetry energy over a broad region ($x$ = 1, 2, 3),
one has to search for other constraints.
Fig.~\ref{v2np} shows predicted neutron and proton elliptic flows in Au + Au reaction
under the FOPI-LAND experimental conditions (here we use the data with b = 7.2 fm case) and geometry \cite{cozma13}. The experimental data
was multiplied by a factor $1.15$ owing to dispersion of the reaction plane \cite{cozma15}.
From Fig.~\ref{v2np} (a), it is seen that our model give somewhat lower value of the elliptic flow of neutrons.
However, from Fig.~\ref{v2np} (b), it is seen that the predicted elliptic flow of protons fit experimental data quite well. In our model, free nucleons are identified by their local densities $\rho_{free} \leq \rho_{0}/8$, which corresponding to deuteron's nucleon average density $0.02 fm^{-3}$. The identification standard of free nucleons affects the value of $V_{2}^{n}$ and $V_{2}^{p}$, but does not affect the ratio of $V_{2}^{n}/V_{2}^{p}$ much.
With stiffer symmetry energy, the value of nucleon elliptic flow should become larger. However, this trend seems not right for the stiffer symmetry energy $x$ = 0. This abnormal behavior may be caused by the competing effects of the SRC and the symmetry energy and deserve further study. From Fig.~\ref{v2np}, at such experimental conditions and geometry, effects of the symmetry energy on both proton and neutron elliptic flows can not be seen clearly. One way to enlarge the effects of the symmetry energy is the relative changes of proton and neutron elliptic flows, such as the ratio of neutron and proton elliptic flows $V_{2}^{n}/V_{2}^{p}$. From this figure, one can see that without SRC, values of both neutron elliptic flow and proton elliptic flow are larger than those with SRC. This is understandable since the SRC of nucleons decreases the anisotropic emissions of neutrons and protons.

\begin{figure}[th]
\centering
\includegraphics[width=0.5\textwidth]{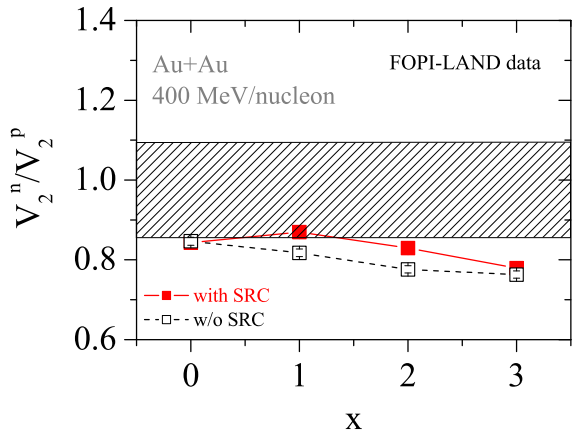}
\caption{(Color online) Same as Fig.~\ref{v2np}, but for the ratio of $V_{2}^{n}/V_{2}^{p}$. The effects of the SRC of nucleon-nucleon on the $V_{2}^{n}/V_{2}^{p}$ with same $x$ parameters are also shown.}
\label{v2rnp}
\end{figure}
Shown in Fig.~\ref{v2rnp} is predicted elliptic flow
ratios of neutron and proton $V_{2}^{n}/V_{2}^{p}$ with different symmetry energies as well as
experimental data \cite{cozma13}. Since stiffer symmetry energy/symmetry potential
causes relatively more neutrons to emit in the direction
perpendicular to the reaction plane \cite{yong07}, one sees larger values of
elliptic flow ratios of neutron and proton
$V_{2}^{n}/V_{2}^{p}$ with stiffer symmetry energies.
With the SRC of nucleon-nucleon in the transport model, values of the $V_{2}^{n}/V_{2}^{p}$ ratio
are larger than that without the SRC of nucleon-nucleon. This is because the SRC of nucleon-nucleon
cause neutron and proton to be correlated together, the value of $V_{2}^{n}/V_{2}^{p}$ ratio trends to
unity. Owing to the competing effects of the SRC and the symmetry energy, for $x$ = 0 case,
the effects of symmetry energy on the trend of the ratio of $V_{2}^{n}/V_{2}^{p}$ with the SRC changes compared with that without the SRC. 

On the whole, the sensitivity of the observable $V_{2}^{n}/V_{2}^{p}$ to the symmetry enegy at FOPI-LAND experimental conditions and geometry is smaller than that of the FOPI $\pi^{-}/\pi^{+}$ ratio. Other nucleon observables should be further explored.

Fig.~\ref{v2rnp}
indicates the FOPI-LAND elliptic flow experimental data does not favor very
soft symmetry energy ($x$ = 2, 3). Combining the studies of nucleon elliptic flow and previous $\pi^{-}/\pi^{+}$ ratio,
one may roughly obtain the symmetry energy stiffness parameter $x$ = 1. It in fact
corresponds a mildly soft density-dependent symmetry energy at supra-saturation densities.
While the specific density region of the present constraints on the
nuclear symmetry energy needs to be further studied \cite{liuy2015}.

The small effects of the symmetry energy on pionic and nucleonic observables, pion suffering from unclear $\pi-N-\Delta$ dynamics and pion in-medium effect, neutron detection efficiency, bound or unbound nucleon identifications, nucleon in-medium and isospin strong interactions as well as all kinds of experimental measurement errors, etc., all affect the probe of the symmetry energy, not to mention uncertainties and complexities of nuclear transport models, thus fully convincing constraints of the symmetry energy at high-density are not easy to achieve.

It is, however, interesting to see that the present result on the symmetry energy stiffness parameter $x$ = 1 (which corresponds a slope of symmetry energy at saturation density ($L (\rho_{0}) \equiv 3\rho_{0}dEsym(\rho)/d\rho$ = 37 MeV) agrees with the recent result $L (\rho_{0})$ = 20 -- 66 MeV quite well by comparing the available data on the electric dipole polarizability with the predictions of the random-phase
approximation, using a representative set of nuclear energy density functionals \cite{edit15}.

\section{Conclusions}

In summary, by incorporating the short-range correlations of nucleon-nucleon
and the in-medium inelastic baryon-baryon scattering cross section into the isospin-dependent
transport model and based on the FOPI and FOPI-LAND experimental measurements, I cross-checked the $\pi^{-}/\pi^{+}$ ratio and the ratio of neutron elliptic flow and
proton elliptic flow $V_{2}^{n}/V_{2}^{p}$ in Au+Au collision.
A mildly soft symmetry energy at supra-saturation densities supports both FOPI and FOPI-LAND experimental measurements.
The studies also show that both the short-range correlations of nucleon-nucleon and the in-medium inelastic baryon-baryon cross section
play important role in probing the symmetry energy with heavy-ion collisions.

Since the symmetry energy plays
crucial roles in both nuclear physics and astrophysics, more subjects, such as the density region that some observables probed and more sensitive observables to the symmetry energy at high densities, deserve further study.

\section*{Acknowledgements}
The author thanks M. D. Cozma for providing the FOPI-LAND elliptic
flow analysis routine and helpful discussions. The work was carried out at National
Supercomputer Center in Tianjin, and the calculations were
performed on TianHe-1A. The work is supported by the National
Natural Science Foundation of China under Grant Nos. 11375239,
11435014.

\end{document}